\documentclass[12pt,amsfonts]{article}
\begin{document}
\def\p {{\partial}}
\def\n {{\nu}}
\def\m {{\mu}}
\def\a {{\alpha}}
\def\bt {{\beta}}
\def\f {{\phi}}
\def\th {{\theta}}
\def\g {{\gamma}}
\def\eps {{\epsilon}}
\def\e {{\psi}}
\def\la {{\lambda}}
\def\na {{\nabla}}
\def\k {\chi}
\def\bn {\begin{eqnarray}}
\def\en {\end{eqnarray}}
\title{Hamiltonian formulation of systems with linear velocities within Riemann-Liouville
fractional derivatives}
\maketitle
\begin{center}
\author{\textbf{Sami I. Muslih}
\vspace{0.2cm}\footnote{E-mail:
smuslih@ictp.trieste.it}\\Department
of Physics, Al-Azhar University, Gaza, Palestine\\and\\
International Center for Theoretical Physics(ICTP),\\Trieste,
Italy\\
\vspace{1cm} \textbf{Dumitru Baleanu} \vspace{0.2cm}\footnote{On
leave of absence from Institute of Space Sciences, P.O.BOX, MG-23,
R 76900, Magurele-Bucharest, Romania,E-mails:
dumitru@cankaya.edu.tr,
baleanu@venus.nipne.ro}\\
Department of Mathematics and Computer Sciences, Faculty of Arts
and Sciences,
\c{C}ankaya University- 06530, Ankara, Turkey }\\
\end{center}
\hskip 5 cm
\begin{abstract}
 The link between the treatments of constrained systems with fractional derivatives by
 using both  Hamiltonian and
 Lagrangian formulations is studied.
 It is shown that both treatments for systems with linear velocities are
 equivalent.
\end{abstract}

\newpage

\section{Introduction}
    The generalization of the concept of derivative and integral to a
non-integer order $\alpha$ has been subjected to several
approaches and some various alternative definitions of fractional
derivatives appeared [1-6]. In the last few years fractional
calculus was applied successfully in various areas , e.g.
chemistry , biology, modelling and identification, electronics and
wave propagation. Fractional calculus , has played an important
role in engineering, science, and pure and applied mathematics
[7-9]. Fractional derivatives were applied  in recent studies of
scaling phenomena [10-12]. Classical mechanics is one of the
fields when fractional calculus found many applications [13-19].
 Riewe  has used the fractional calculus to obtain a formalism which can be
applied for both conservative and non conservative systems [13,
14]. Although many laws of nature can be obtained using certain
functionals and the theory of calculus of variations, not all laws
can be obtained by this manner. As it is known, almost all systems
contain internal damping, yet traditional energy based approach
cannot be used to obtain equations describing the behavior of a
non conservative system [13,14].Using the fractional calculus one
can obtain the Lagrangian and the Hamiltonian equations of motion
for the nonconservative systems .

The understanding of constrained dynamics \cite{20}, both at the
classical and quantum level, has been a subject of long standing
theoretical interest, which has seen important contributions ever
since Dirac's quantization of the electromagnetic field. The path
integral approach and the canonical one are two main approaches of
quantization.

 Recently, an extension of the simplest fractional problem and the
fractional variational problem of Lagrange was obtained [17,18].
Even more recently,this approach was extended to Lagrangians with
linear in velocities [21,22] , which represents a typical example
of second-class constrained systems in Dirac's classification
[20]. These Lagrangians are important because their
Euler-Lagrangian equations become systems of first order
differential equations in contrast with second order corresponding
to the regular ones. In addition, these systems may possess gauge
symmetries and gauge ambiguities.

From these reasons it is interesting to study the fractional
Hamiltonian of constrained systems.

The aim of this paper is to obtain the fractional Hamiltonian
equations of motion for Lagrangians with linear velocities.

The plan of our paper is as follows:

In Sect. 2 some basic tools of fractional derivatives as well as
Riewe's approach of the fractional Lagrangian and Hamiltonians are
presented. In Sect. 3 the Euler-Lagrange equations were obtained
using the Agrawal's approach and the fractional formulation of
systems with constraints is introduced. In Sect 4 the fractional
Hamiltonian of the systems possessing linear velocities is
analyzed. Conclusion is presented in section 5.

\section{Fractional Lagrangian and Hamiltonian formulations}
In this section we briefly present the definition of the left and
right derivatives together with Riewe's formulation of Lagrangian
and Hamiltonian dynamics.
 The~ left~ Riemann-Liouville~ fractional~ derivative is defined
 as

\begin{equation}
{}_a\textbf{D}_t^{\alpha}f(t)=
\frac{1}{\Gamma{(n-\alpha)}}\left(\frac{d}{dt}\right)^{n}\int\limits_a^t(t-\tau)^{n-\alpha-1}f(\tau)d\tau,
\end{equation}

and the~ right ~Riemann-Liouville ~fractional~ derivative has the
 form

\begin{equation}
{}_t\textbf{D}_b^{\alpha}f(t)=
\frac{1}{\Gamma{(n-\alpha)}}\left(-\frac{d}{dt}\right)^{n}\int\limits_t^b(\tau-t)^{n-\alpha-1}f(\tau)d\tau,
\end{equation}
where the order $\alpha$ fulfills $n-1\leq\alpha <n$ and $\Gamma$
represents the Euler's gamma function. If $\alpha$ is an integer,
these derivatives are defined in the usual sense, i.e.,
\begin{equation}
{}_a\textbf{D}_t^{\alpha}f(t)
=\left(\frac{d}{dt}\right)^{\a},~~\large{{}_t\textbf{D}_b^{\alpha}f(t)}
= \left(-\frac{d}{dt}\right)^{\a}, ~\a=1,2,... .
\end{equation}

 Now we shall briefly review Riewe's formulation of
fractional generalization of Lagrangian and Hamiltonian equations
of motion [13,14]. The starting point is the action function of
the form

\begin{equation}
S=\int_{a}^{b} L(\{q_{n}^{r}, Q_{n'}^{r}\}, t)dt.
\end{equation}
Here the generalized coordinates are defined as
\begin{equation}
q_{n}^{r}:= \large{({}_a\textbf{D}_t^{\alpha})^{n}
x_{r}(t)},~Q_{n'}^{r}:= ({}_t\textbf{D}_b^{\alpha})^{n'} x_{r}(t),
\end{equation}
and $r=1,2,..., R $ represents the number of fundamental
coordinates, $ n=0,..., N,$ the sequential order of the
derivatives defining the generalized coordinates $q$, and
$n'=1,..., N'$ the sequential order of the derivatives in
definition of the coordinates $Q$. A necessary condition for $S$
to posses  an extremum for given functions $x_{r}(t)$ is that
$x_{r}(t)$ fulfill  the Euler-Lagrange equations [13,14]
\begin{equation}
\frac{\p L}{\p q_{0}^{r}} +
\sum_{n=1}^{N}\large{({}_t\textbf{D}_b^{\alpha})^{n}}\frac{\p
L}{\p q_{n}^{r}} +
\sum_{n'=1}^{N'}({}_a\textbf{D}_t^{\alpha})^{n'}\frac{\p L}{\p
Q_{n'}^{r}}  =0.
\end{equation}

Using the references [13,14], the generalized momenta have the
following  form \bn && p_{n}^{r}=
\sum_{k=n+1}^{N}\large{({}_t\textbf{D}_b^{\alpha})^{k-n-1}}\frac{\p
L}{\p q_{k}^{r}},\nonumber\\
&& \pi_{n'}^{r}=
\sum_{k=n'+1}^{N'}\large{({}_a\textbf{D}_t^{\alpha})^{k-n'-1}}\frac{\p
L}{\p Q_{k}^{r}}. \en

Thus, the canonical Hamiltonian is given by
\begin{equation}
H = \sum_{r=1}^{R}\sum_{n=0}^{N-1} p_{n}^{r}q_{n+1}^{r} +
\sum_{r=1}^{R}\sum_{n'=0}^{N'-1} \pi_{n'}^{r}Q_{n'+1}^{r} - L.
\end{equation}

The Hamilton's  equations of motion are as follows
\begin{equation}
\frac{\p H}{\p q_{N}^{r}}=0,~\frac{\p H}{\p Q_{N'}^{r}}=0.
\end{equation}
For $n=1,..., N,~n'=1,..., N'$ we have the following equations of
motion

\bn &&\frac{\p H}{\p q_{n}^{r}}= \large{{}_t\textbf{D}_b^{\alpha}
p_{n}^{r}},~~\frac{\p H}{\p Q_{n'}^{r}}=\large{{}_a\textbf{D}_t^{\alpha} \pi_{n'}^{r}},\\
&&\frac{\p H}{\p q_{0}^{r}} = - \frac{\p L}{\p q_{0}^{r}}=
\large{{}_t\textbf{D}_b^{\alpha} p_{0}^{r}} +
\large{{}_a\textbf{D}_t^{\alpha} \pi_{0}^{r}}.\en

The remaining equations are given by \bn &&\frac{\p H}{\p
p_{n}^{r}}=q_{n+1}^{r}= \large{{}_a\textbf{D}_t^{\alpha}
q_{n}^{r}}, ~~\frac{\p H}{\p
\pi_{n'}^{r}}=Q_{n+1}^{r}=\large{{}_t\textbf{D}_b^{\alpha}
Q_{n'}^{r}}, \\
&&\frac{\p H}{\p t} = - \frac{\p L}{\p t}, \en where, $n=0,...,
N,~n'=1,..., N'$.

\section{Fractional Euler-Lagrange equations}

Recently  Agrawal , have obtained the Euler-Lagrange equations for
fractional variational problems [17]. In the following we like to
present briefly  his approach.

Consider the action function
\begin{equation}
S[q_{0}^{1},....q_{0}^{R}] =\int_{a}^{b}L(\{q_{n}^{r},
Q_{n'}^{r}\}, t)dt,
\end{equation}

subject to the independent constraints
\begin{equation}
\Phi_{m}(t, q_{0}^{1},..., q_{0}^{R}, q_{n}^{r},Q_{n'}^{r})=0, ~m
< R,
\end{equation}
where the generalized coordinates are defined as
\begin{equation}
q_{n}^{r}:= \large{({}_a\textbf{D}_t^{\alpha})^{n}
x_{r}(t)},~Q_{n'}^{r}:= \large{({}_t\textbf{D}_b^{\bt})^{n'}
x_{r}(t)}.
\end{equation}

Then, the necessary condition for the curves $q_{0}^{1},....,
q_{0}^{R}$ with the boundary conditions
$q_{0}^{r}(a)=q_{0}^{ra},~q_{0}^{r}(b)=q_{0}^{rb},r=1, 2,,...,R$,
to be an extremal of the functional given by equation (14) is that
the functions $q_{0}^{r}$ satisfy the following Euler-Lagrange
equations [17]:
\begin{equation}
\frac{\p \bar{L}}{\p q_{0}^{r}} +
\sum_{n=1}^{N}\large{({}_t\textbf{D}_b^{\alpha})^{n}}\frac{\p \bar
{L}}{\p q_{n}^{r}} +
\sum_{n'=1}^{N'}\large{({}_a\textbf{D}_t^{\a})^{n'}}\frac{\p
\bar{L}}{\p Q_{n'}^{r}}  =0,
\end{equation}

where $\bar{L}$ has the form [17]
\begin{equation}
\bar{L}(\{q_{n}^{r}, Q_{n'}^{r}\}, t, \lambda_{m}(t))=
L(\{q_{n}^{r}, Q_{n'}^{r}\}, t) + \lambda_{m}(t)\Phi_{m}(t,
q_{0}^{1},..., q_{0}^{R}, q_{n}^{r},Q_{n'}^{r}).
\end{equation}
Here the multiple $\lambda_{m}(t)\in R^{m}$ are continuous on $[a,
b]$.

\subsection{Fractional Hamiltonian formulation of Agrawal's
approach}

In order to obtain the Hamilton's equations for the the fractional
variational problems, we re-define the left and the right
canonical momenta as : \bn\label{bobo} && p_{n}^{r}=
\sum_{k=n+1}^{N}\large{({}_t\textbf{D}_b^{\alpha})^{k-n-1}}\frac{\p
\bar{L}}{\p q_{k}^{r}},\nonumber\\
&& \pi_{n'}^{r}=
\sum_{k=n'+1}^{N'}\large{({}_a\textbf{D}_t^{\a})^{k-n'-1}}\frac{\p
\bar{L}}{\p Q_{k}^{r}}. \en

Using (\ref{bobo}),the canonical Hamiltonian becomes
\begin{equation}
{\bar{H}} = \sum_{r=1}^{R}\sum_{n=0}^{N-1} p_{n}^{r}q_{n+1}^{r} +
\sum_{r=1}^{R}\sum_{n'=0}^{N'-1} \pi_{n'}^{r}Q_{n'+1}^{r} -
{\bar{L}}.
\end{equation}
Then, the modified canonical equations of motion are obtained as
\bn&&\{q_{n}^{r}, \bar{H}\}= \large{{}_t\textbf{D}_b^{\alpha}
p_{n}^{r}},~\{Q_{n'}^{r}, \bar{H}\} =\large{{}_a\textbf{D}_t^{\alpha} \pi_{n'}^{r}},\\
&&\{q_{0}^{r}, \bar{H}\} = \large{{}_t\textbf{D}_b^{\alpha}
p_{0}^{r}} + \large{{}_a\textbf{D}_t^{\alpha} \pi_{0}^{r}},\en
where, $n=1,..., N,~n'=1,..., N'$.

The other set of equations of motion are given by \bn &&
\{p_{n}^{r}, \bar{H}\}
=q_{n+1}^{r}=\large{{}_a\textbf{D}_t^{\alpha}
q_{n}^{r}},~\{\pi_{n'}^{r}, \bar{H}\}= Q_{n+1}^{r}=
\large{{}_t\textbf{D}_b^{\alpha}
Q_{n'}^{r}}, \\
&&\frac{\p \bar{H}}{\p t} = - \frac{\p \bar{L}}{\p t}. \en Here,
$n=0,..., N,~n'=1,..., N'$ and the commutator $\{, \}$ is the
Poisson's bracket  defined as
\begin{equation}
\{A, B\}_{q_{n}^{r}, p_{n}^{r}, Q_{n'}^{r}, \pi_{n'}^{r}}
=\frac{\p A}{\p q_{n}^{r}}\frac{\p B}{\p p_{n}^{r}}- \frac{\p
B}{\p q_{n}^{r}}\frac{\p A}{\p p_{n}^{r}} + \frac{\p A}{\p
Q_{n'}^{r}}\frac{\p B}{\p \pi_{n'}^{r}}- \frac{\p B}{\p
Q_{n'}^{r}}\frac{\p A}{\p \pi_{n'}^{r}},
\end{equation}
where, $n=0,..., N,~n'=1,..., N'$.

\section{Equivalence of fractional Hamiltonian and Lagrangian formulations for systems with
linear velocities}

Recently, for $0<\alpha\leq 1$, the Lagrangians with linear
velocities were investigated in [21]. For example, the
Euler-Lagrange equations of the following Lagrangian

\begin{equation}
L'=a_{j}\left(q^{i}\right){}_a \textbf{D}_t^{\alpha}
q^{j}-V\left(q^{i}\right) \ ,
\end{equation}
 were obtained as [21]

\begin{equation}
\frac{\partial a_{j}(q^i)}{\partial q^{k}}{}_a
\textbf{D}_t^{\alpha} q^{j}+{}_t
\textbf{D}_b^{\alpha}a_{k}\left(q^{i}\right)-\frac{\partial
V\left(q^{i}\right)}{\partial q^{k}}=0 \ .
\end{equation}

Now we would like to obtain the Hamiltonian equations of motion
for the same model. Let us define
\begin{equation}
x_{j}^{n}= ({}_a \textbf{D}_t^{\alpha})^{n} q_{j},~~~n=0,1,...,
N-1,~~~j= 1,2, ...,R.
\end{equation}
The generalized momenta are given by,
\begin{equation}
p_{j}^{0} = \frac{\p L}{\p x_{j}^{0}}= a_{j}(x_{i}^{0}),~~~,
p_{j}^{1} = \frac{\p L}{\p x_{j}^{1}}= 0.
\end{equation}

The canonical Hamiltonian reads as
\begin{equation}
H = \left(p_{j}^{0} - a_{j}(x_{i}^{0})\right)x_{j}^{1} +
V(x_{i}^{0}).
\end{equation}

The Hamiltonian equations of motion are calculated as
\begin{equation}
\frac{\p H}{\p x_{j}^{1}}= p_{j}^{0} - a_{j}(x_{i}^{0})= ({}_t
\textbf{D}_b^{\alpha})p_{j}^{1}=0.
\end{equation}
In fact, this equation is the primary constraint in the Dirac's
formalism. The other equations of motion are calculated as \bn &&
\frac{\p H}{\p x_{k}^{0}}= -\frac{\partial
a_{j}(x_{i}^{0})}{\partial x_{k}^{0}} x_{j}^{1} + \frac{\partial
V\left(x_{i}^{0}\right)}{\partial q^{k}} = ({}_t
\textbf{D}_b^{\alpha})p_{k}^{0},\\
&&\frac{\p H}{\p p_{k}^{0}}= x_{k}^{1}= ({}_a
\textbf{D}_t^{\alpha}) q_{k} \en

Making use of equation (31) we obtain
\begin{equation}
\frac{\partial a_{j}(q^i)}{\partial q^{k}}{}_a
\textbf{D}_t^{\alpha} q^{j}+{}_t
\textbf{D}_b^{\alpha}a_{k}\left(q^{i}\right)-\frac{\partial
V\left(q^{i}\right)}{\partial q^{k}}=0 \ .
\end{equation}

It is obvious the equivalence between equations (34) and (27). An
interesting point to be specified here is that, primary
constraints in the Dirac's formalism are present as equations of
motion in our treatment, while the other Hamiltonian equations of
motion are equivalent to the Lagrangian equations of motion as
given in [21].

\section{Conclusion}
One of the main problems encountered in applying the fractional
calculus to a given singular Lagrangian and Hamiltonian is the
existence of multiple choices of the possible fractional
generalizations . In addition, the solutions of the fractional
Euler-Lagrange equations contain more information than the
classical ones.
 In this paper, Hamiltonian
equations have been obtained for systems with linear velocities,
in the same manner as those obtained by using the formulation of
Euler-lagrange equations for variational problems introduced by
one of us [21]. We study the general model  for systems with
linear velocities and it was observed that the Hamiltonian and
Lagrangian equations which are obtained by the two methods are in
exact agreement.

\section*{Acknowledgments}

S. M. would like to thank the Abdus Salam International Center for
Theoretical Physics, Trieste, Italy, for support and hospitality
during the preparation of this work. This work was done within the
framework of the Associateship Scheme of the Abdus Salam ICTP.\\
D. B.  would like to thank  O. Agrawal and J. A. Tenreiro Machado
for interesting discussions. This work is partially supported by
the Scientific and Technical Research Council of Turkey.

\end{document}